\documentclass[a4paper,11pt]{article}

\usepackage{amsmath,amssymb,mathtools,amsthm}
\usepackage[shortlabels]{enumitem}
\usepackage[colorlinks=true,allcolors=blue]{hyperref}
\usepackage[margin=1.8cm]{geometry}
\usepackage[breakable]{tcolorbox}
\usepackage{tgpagella,eulervm}
\definecolor{lightgray}{rgb}{0.9,0.9,0.9}
\definecolor{lightblue}{RGB}{209,228,245}%{236,247,254}
\usepackage[font=footnotesize]{caption}
\usepackage[sort&compress,numbers]{natbib}
\newcommand{\sgn}{\operatorname{sgn}}
\newtheorem{definition}{Definition}

\newtheorem{remark}{Remark}
\newtheorem{algorithm}{Algorithm}

\usepackage{abstract}

\title{Stochastic simulations of nonlinear reaction-diffusion\\ equations using an exponential integrator}

\author{Elliot J. Carr\footnote{Corresponding Author (\href{elliot.carr@qut.edu.au}{elliot.carr@qut.edu.au})}\\ \small School of Mathematical Sciences, Queensland University of Technology (QUT), Brisbane, Australia}
\date{\medskip}

\newcommand{\algbox}[1]{\small\vspace{0.0cm}\fcolorbox{lightblue}{lightblue}{\parbox{0.99\textwidth}{\color{black}\vspace{-0.2cm}#1 \vspace{-0.2cm}}}\vspace{0.2cm}}

\begin{document}
\maketitle
\vspace*{-1cm}
\begin{abstract} 
Stochastic simulations can be generated from deterministic reaction-diffusion equations by discretising in space and time and interpreting coefficients in the resulting system of discretised equations as probabilities governing movement and reaction events. In this paper, we present a novel variant of this approach for nonlinear reaction-diffusion equations that employs an exponential integrator when discretising in time. The proposed method yields valid probabilities, defined by the entries of appropriate matrix functions, without the strict conditions on the time step required by a commonly-employed time discretisation scheme. Simulation results presented for one and two dimensional Porous-Fisher type models demonstrate the veracity of the method across several test problems.
\end{abstract}


\section{Introduction}
Reaction-diffusion models are fundamental in applied mathematics and widely used in biological and ecological applications. Arguably, the most famous reaction-diffusion model is Fisher-KPP \cite{murray_2002}, a semi-linear model involving a linear diffusion term and a nonlinear reaction term. The Fisher-KPP model assumes the rate of reaction decreases linearly to zero as the population density increases to the carrying capacity but that the rate of diffusion (diffusivity) is unaffected by such changes in the population density. To address this limitation, several extensions of the Fisher-KPP model have been proposed that include a nonlinear diffusion term, where the diffusivity either increases or decreases with increasing population density \cite{cai_2006}, that is, individuals are either assumed to be less motile due to overcrowding (e.g. restricted movement) or more motile due to overcrowding (e.g. restricted resources). Notably, much attention has explored nonlinear degenerate diffusion \cite{sherratt_1996}, where the diffusion term vanishes (diffusivity is zero) at zero population density; a property that is commonly achieved by combining the power law diffusivity of the porous medium equation \cite{vazquez_2006} with the Fisher-KPP model to give the so-called Porous-Fisher model \cite{witelski_1995,simpson_2024,sherratt_1996,mccue_2019,buenzli_2020,crossley_2026,depablo_1998,ichida_2025,stokes_2024}. In contrast to Fisher-KPP, the Porous-Fisher model allows for solutions with compact support and sharp invading fronts into regions of zero population density, which is desirable from a mathematical modelling standpoint as such features are consistent with observations in many ecological and biological applications (e.g. wound healing, tumour invasion, animal dispersal) \cite{murray_2002}.

Reaction-diffusion models are commonly classified as either \textit{deterministic} or \textit{stochastic}. Deterministic models typically consider a continuous population density with a deterministic partial differential equation (PDE) governing movement and reaction, while stochastic models typically consider discrete individuals with designated probabilities governing movement and reaction events. An ongoing problem in this area of research is to establish connections between deterministic and stochastic models of reaction-diffusion \cite{gavagnin_2018}. One way forward is to start with a stochastic model for individual movement and reaction and derive a deterministic PDE model for the population density in the continuum limit \cite{codling_2008}. An alternative, less common strategy, and the focus of this work, is to start with the deterministic PDE model and discretise in space and time to derive a corresponding stochastic model \cite{codling_2008}. Here, the standard approach is to use a simple forward Euler discretisation in time \cite{cai_2006,anderson_1998,carwood_2026,codling_2008,vigelius_2012}, which leads to strict conditions on the time step to ensure probabilities in the stochastic model remain in $[0,1]$ \cite{carr_2025}. In this paper, we outline an alternative method for deriving such stochastic models that employs an exponential integrator for discretising in time, a class of explicit methods that have favourable properties and that typically allow for larger time steps than traditional explicit methods \cite{hochbruck_2010,minchev_2005}. In short, our novel method yields probabilities governing movement and reaction events, expressed in terms of the entries of a matrix exponential and a closely-related matrix phi function, that are uninhibited by any restrictions on the time step to ensure probabilities remain in $[0,1]$. The work builds on a recent paper concerning linear and semi-linear problems \cite{carr_2026} with simulation results presented for 1D and 2D Porous-Fisher type models demonstrating good agreement with the corresponding deterministic model. 

\begin{definition}\label{def:1}
Let $\mathbf{A}\in\mathbb{R}^{n\times n}$, $\mathbf{b}\in\mathbb{R}^{n}$ and $\psi$ be a matrix function such that $\psi(\mathbf{A})\in\mathbb{R}^{n\times n}$. Throughout this paper we adopt the following notation and terminology. 
\begin{enumerate}[(i),itemsep=0pt,topsep=0pt]
\item $[\mathbf{A}]_{ij}$ or $[\mathbf{A}]_{i,j}$ denotes the entry of $\mathbf{A}$ in the $i$th row and $j$th column. 
\item $[\mathbf{b}]_{i}$ denotes the entry of $\mathbf{b}$ in the $i$th row.
\item $[\psi(\mathbf{A})]_{ij}$ denotes the entry of $\psi(\mathbf{A})$ in the $i$th row and $j$th column.
\item $\mathbf{A}$ is \textit{essentially nonnegative} if $[\mathbf{A}]_{ij}\geq 0$ for all $i\neq j$.
\item $\psi(\mathbf{A})$ is \textit{nonnegative} if $[\psi(\mathbf{A})]_{ij}\geq 0$ for all $i=1,\hdots,n$ and $j=1,\hdots,n$.
\item $\mathbf{A}$ is a \textit{column transition-rate matrix} if $\mathbf{A}$ is \textit{essentially nonnegative} and $\sum_{i=1}^{n}[\mathbf{A}]_{ij} = 0$ for all $j = 1,\hdots,n$.
\item $\psi(\mathbf{A})$ is a \textit{column stochastic matrix} if $\psi(\mathbf{A})$ is nonnegative and $\sum_{i=1}^{n}[\psi(\mathbf{A})]_{ij} = 1$ for all $j = 1,\hdots,n$.
\end{enumerate}
\end{definition}

\section{Stochastic method}
\label{sec:stochastic_framework}
Our method begins with a standard deterministic PDE model of nonlinear reaction-diffusion, which we then discretise in space and time to derive a corresponding stochastic model. As we will see later in sections \ref{sec:1D} and \ref{sec:2D}, carrying out the discretisation in space yields a system of nonlinear differential equations that can be expressed in the form:
\begin{gather}
\label{eq:ode_nonlinear}
\frac{\text{d}\mathbf{u}}{\text{d}t} = \mathbf{g}(\mathbf{u}) + \mathbf{f}(\mathbf{u}),
\end{gather}
with initial condition $\mathbf{u}(0) = \mathbf{u}_{0}$. Here, $\mathbf{u}(t)\in\mathbb{R}^{n}$ is the solution/state vector, $n$ is the number of discrete unknowns/nodes, and $\mathbf{g}(\mathbf{u})$ and $\mathbf{f}(\mathbf{u})$ are vector-valued functions containing the spatially-discretised diffusion and reaction terms, respectively. 

The novelty of our method lies in the use of an exponential integrator to discretise (\ref{eq:ode_nonlinear}) in time. Here, we obtain approximations to $\mathbf{u}(t)$ at the discrete times $t_{k} = k\tau$ for $k = 0,1\hdots,M$, where $\tau = T/M>0$ is the time step and $M$ is the number of time steps. This is achieved by approximating the nonlinear system (\ref{eq:ode_nonlinear}) over the time interval $[t_{k},t_{k+1}]$ by the linearised system
\begin{gather}
\label{eq:ode_linearised}
\frac{\text{d}\mathbf{u}}{\text{d}t} = \mathbf{A}_{k}\mathbf{u} + \mathbf{b}(\mathbf{u}_{k}),
\end{gather}
where $\mathbf{b}(\mathbf{u}_{k}) = \mathbf{g}(\mathbf{u}_{k}) + \mathbf{f}(\mathbf{u}_{k}) - \mathbf{A}(\mathbf{u}_{k})\mathbf{u}_{k}$ and $\mathbf{A}_{k} = \mathbf{A}(\mathbf{u}_{k})$ is the Jacobian matrix of $\mathbf{g}(\mathbf{u})$ evaluated at $\mathbf{u}=\mathbf{u}_{k}$. Essentially, the nonlinear system (\ref{eq:ode_nonlinear}) is linearised using the Taylor approximations $\mathbf{g}(\mathbf{u}) \approx \mathbf{g}(\mathbf{u}_{k}) + \mathbf{A}_{k}(\mathbf{u}-\mathbf{u}_{k})$ and $\mathbf{f}(\mathbf{u}) \approx \mathbf{f}(\mathbf{u}_{k})$ to produce a semi-linear system (\ref{eq:ode_linearised}) that can be solved exactly using the integrating factor $e^{t\mathbf{A}_{k}}$. This procedure ultimately yields the exponential integrator \cite{hochbruck_2010,minchev_2005}
\begin{gather}
\label{eq:exponential_integrator}
\mathbf{u}_{k+1} = e^{\tau\mathbf{A}_{k}}\mathbf{u}_{k} + \tau\varphi(\tau\mathbf{A}_{k})\mathbf{b}(\mathbf{u}_{k}),
\end{gather}
which is then applied for $k = 0,\hdots,M-1$ to compute approximations $\mathbf{u}_{k}\approx\mathbf{u}(t_{k})$ ($k = 1,\hdots,M$) of the nonlinear system (\ref{eq:ode_nonlinear}). Note that (\ref{eq:exponential_integrator}) is explicit and involves two matrix functions, the matrix exponential and one of the so-called matrix phi-functions \cite{higham_2008,skaflestad_2009,al-mohy_2026}, which can be defined by power series that converge for all $\mathbf{A}\in\mathbb{R}^{n\times n}$: $e^{\mathbf{A}} = \sum_{i=0}^{\infty} \mathbf{A}^{i}/i!$ and $\varphi(\mathbf{A}) = \sum_{i=0}^{\infty} \mathbf{A}^{i}/(i+1)!$ \cite{schmelzer_2008,wu_2018}. Moreover, as evident from its expansion, $\smash{\mathbf{u}_{k+1} = \mathbf{u}_{k}+\tau[\mathbf{g}(\mathbf{u}_{k}) + \mathbf{f}(\mathbf{u}_{k})] + \frac{\tau^{2}}{2}\mathbf{A}_{k}[\mathbf{g}(\mathbf{u}_{k}) + \mathbf{f}(\mathbf{u}_{k})] + O(\tau^{3})}$, the exponential integrator (\ref{eq:exponential_integrator}) is a first order method (local error order two) if $\mathbf{f}(\mathbf{u})\neq \mathbf{0}$ and a second order method (local error order three) if $\mathbf{f}(\mathbf{u}) = \mathbf{0}$ (i.e. pure diffusion problem). 

The key observation now is the following: If the Jacobian matrix $\mathbf{A}_{k}$ is a column transition-rate matrix (Definition \ref{def:1}(vi)), then $e^{\tau\mathbf{A}_{k}}$ and $\varphi(\tau\mathbf{A}_{k})$ are column stochastic matrices (Definition \ref{def:1}(vii)) for all $\tau>0$ \cite{carr_2026}. In this case, the exponential integrator (\ref{eq:exponential_integrator}) can be used to generate discrete-time stochastic realisations of the underlying deterministic solution $\mathbf{u}_{k}$ ($k = 1,\hdots,M$). The general strategy, developed in our previous work \cite{carr_2026}, is to assume that $[\mathbf{u}_{k}]_{i}$ comprises $N([\mathbf{u}_{k}]_{i})$ individual discrete units that act independently. Here, $N(\mu) = \lceil|N_{s}\mu|\rceil$ is the smallest nonnegative integer greater than $|N_{s}\mu|$ with $N_{s}$ specifying the number of individual discrete units when $\mu = 1$. The entries of $e^{\tau\mathbf{A}_{k}}$ and $\varphi(\tau\mathbf{A}_{k})$ then provide probabilities governing movement and reaction events across the $n$ spatial nodes. The resulting stochastic model is described in Algorithm \ref{alg:stochastic_model}, where $\widehat{\mathbf{u}}_{k}$ represents a stochastic realisation of $\mathbf{u}_{k}$, $[e^{\tau\mathbf{A}_{k}}]_{ij}$ is the probability that $[\widehat{\mathbf{u}}_{k}]_{j}$ contributes $1/N([\widehat{\mathbf{u}}_{k}]_{j})$ of its value to $[\widehat{\mathbf{u}}_{k+1}]_{i}$ and $[\varphi(\tau\mathbf{A}_{k})]_{ij}$ is the probability that $[\tau\mathbf{b}(\widehat{\mathbf{u}}_{k})]_{j}$ contributes $1/N(\tau\mathbf{b}(\widehat{\mathbf{u}}_{k})]_{j})$ of its value to $[\widehat{\mathbf{u}}_{k+1}]_{i}$. 

\vspace*{-0.2cm}
\noindent
\begin{figure}[h]
\algbox{
\begin{algorithm}[Stochastic model]\mbox{}\\
\label{alg:stochastic_model}
\emph{
\begin{tabular}{@{}l}
Input vector $\mathbf{u}_{0}$, vector functions $\mathbf{g}(\mathbf{u})$ and $\mathbf{f}(\mathbf{u})$ and matrix function $\mathbf{A}(\mathbf{u})$\\
Choose stochastic model discretisation parameter $N_{s}$ and initialise $\widehat{\mathbf{u}}_{0}=\mathbf{u}_{0}$\\
Choose time duration $T$ and number of time steps $M$, and compute time step $\tau = T/M$\\
\textbf{for} $k = 0,\hdots,M-1$\\
\qquad Initialise $\widehat{\mathbf{u}}_{k+1}=\mathbf{0}$\\ 
\qquad Compute $\mathbf{A}_{k} = \mathbf{A}(\widehat{\mathbf{u}}_{k})$ and $\mathbf{b}(\widehat{\mathbf{u}}_{k}) = \mathbf{g}(\widehat{\mathbf{u}}_{k}) + \mathbf{f}(\widehat{\mathbf{u}}_{k}) - \mathbf{A}(\widehat{\mathbf{u}}_{k})\widehat{\mathbf{u}}_{k}$\\
\qquad Compute $e^{\tau\mathbf{A}_{k}}$ and $\varphi(\tau\mathbf{A}_{k})$ \\
\qquad Compute cumulative probabilities $[\mathbf{P}_{0}]_{0,j} = 0$ and $[\mathbf{P}_{0}]_{ij} = \sum_{m=1}^{i}[e^{\tau\mathbf{A}_{k}}]_{mj}$ for all $j=1,\hdots,n$\\
\qquad Compute cumulative probabilities $[\mathbf{P}_{1}]_{0,j} = 0$ and $[\mathbf{P}_{1}]_{ij} = \sum_{m=1}^{i}[\varphi(\tau\mathbf{A}_{k})]_{mj}$ for all $j=1,\hdots,n$\\
\qquad\textbf{for} $\ell = 0,1$\\
\qquad\qquad Set $\mathbf{v} = \widehat{\mathbf{u}}_{k}$ if $\ell = 0$ or $\mathbf{v} = \tau\mathbf{b}(\widehat{\mathbf{u}}_{k})$ if $\ell = 1$\\
\qquad\qquad\textbf{for} $j = 1,\hdots,n$\\
\qquad\qquad\qquad Set $N([\mathbf{v}]_{j}) = \sgn([\mathbf{v}]_{j})\lceil|N_{s}[\mathbf{v}]_{j}|\rceil$\\ 
\qquad\qquad\qquad Generate random number $r\sim U(0,1)$\\
\qquad\qquad\qquad Find $i$ such that $[\mathbf{P}_{\ell}]_{i-1,j} < r < [\mathbf{P}_{\ell}]_{ij}$\\
\qquad\qquad\qquad $[\widehat{\mathbf{u}}_{k+1}]_{i} = [\widehat{\mathbf{u}}_{k+1}]_{i} + [\mathbf{v}]_{j}/N([\mathbf{v}]_{j})$\\
\qquad\qquad\textbf{end}\\
\qquad\textbf{end}\\
\textbf{end}\\
Return $\widehat{\mathbf{u}}_{k}$ for $k=1,\hdots,M$
\end{tabular}}
\end{algorithm}}
\end{figure}
\vspace*{-0.1cm}

In the next sections, we consider a standard deterministic reaction-diffusion model, involving nonlinear diffusivity $D(c)$ and reaction $R(c)$ functions, and then discretise the model in space using a finite volume method to produce the nonlinear ODE system (\ref{eq:ode_nonlinear}). In particular, careful attention is paid to the discretisation of the diffusion term in order to produce a Jacobian matrix satisfying the properties of a \textit{column transition-rate matrix} (Definition \ref{def:1}(vi)).

\section{One dimensional formulation}
\label{sec:1D}
Consider the one dimensional model
\begin{gather}
\label{eq:1D_pde}
\frac{\partial c}{\partial t} = \frac{\partial}{\partial x}\left(D(c)\frac{\partial c}{\partial x}\right) + R(c),\\
\label{eq:1D_IBCs}
c(x,0) = F(x),\quad q(0,t) = 0,\quad q(L,t) = 0,
\end{gather}
where $q(x,t) = D(c(x,t))\frac{\partial c}{\partial x}(x,t)$. Discretisation in space is carried out on a uniform grid consisting of $n$ nodes, $x_{i} = (i-1)h$ for $i = 1,\hdots,n$, and $n$ finite volumes, $\Omega_{i} = [x_{i}^{-},x_{i}^{+}]$ for $i = 1,\hdots,n$, where $h = L/(n-1)>0$, $x_{i}^{-}=\max(0,x_{i}-h/2)$ and $x_{i}^{+} = \min(x_{i}+h/2,L)$. Applying a standard finite volume method by integrating the governing PDE (\ref{eq:1D_pde}) over the finite volumes, inserting the boundary conditions, and employing standard central difference approximations for the diffusion term yields the ODE system:
\begin{align}
\label{eq:1D_ode1}
\frac{\text{d}c_{i}}{\text{d}t} &= \frac{\overline{D}(c_{i},c_{i+1})(c_{i+1}-c_{i})}{V_{i}h} + R(c_{i}),\quad i = 1,\\
\frac{\text{d}c_{i}}{\text{d}t} &= \frac{\overline{D}(c_{i},c_{i+1})(c_{i+1}-c_{i}) - \overline{D}(c_{i-1},c_{i})(c_{i}-c_{i-1})}{V_{i}h}+ R(c_{i}),\quad i = 2,\hdots,n-1,\\
\label{eq:1D_oden}
\frac{\text{d}c_{i}}{\text{d}t} &= \frac{\overline{D}(c_{i-1},c_{i})(c_{i-1}-c_{i})}{V_{i}h}+ R(c_{i}),\quad i = n.
\end{align}
with initial condition $c_{i}(0) = F(x_{i})$, where $c_{i}\approx c(x_{i},t)$, $V_{i} = x_{i}^{+}-x_{i}^{-}$ and $\overline{D}(c_{i},c_{j})$ denotes the inter-nodal approximation of $D(c)$ at $x = (x_{i}+x_{j})/2$. For succinctness of the analysis that follows, we express the ODE system (\ref{eq:1D_ode1})--(\ref{eq:1D_oden}) compactly in the form:
\begin{align*}
\frac{\text{d}c_{i}}{\text{d}t} &= \frac{1}{V_{i}}\sum_{k\in S_{i}}\left[\overline{D}(c_{i},c_{k})\frac{(c_{k}-c_{i})}{h}\right] + R(c_{i}),\quad i = 1,\hdots,n,
\end{align*}
where $S_{i}$ denotes the set of nodes connected to node $i$, that is, $S_{1} = \{2\}$, $S_{i} = \{i-1,i+1\}$ for $i = 2,\hdots,n-1$, and $S_{n} = \{n-1\}$. Equivalently, in vector form we have:
\begin{gather}
\label{eq:1D_ode_c}
\frac{\text{d}\mathbf{c}}{\text{d}t} = \widetilde{\mathbf{g}}(\mathbf{c}) + \widetilde{\mathbf{f}}(\mathbf{c}),
\end{gather}
with initial condition $\mathbf{c}(0) = \mathbf{c}_{0}$, where $\mathbf{c} = (c_{1},\hdots,c_{n})^{T}$, $\mathbf{c}_{0} = (F(x_{1}),\hdots,F(x_{n}))^{T}$, $\mathbf{g}(\mathbf{c}) = (g_{1}(\mathbf{c}),\hdots,g_{n}(\mathbf{c}))^{T}$ and $\mathbf{f}(\mathbf{c}) = (f_{1}(\mathbf{c}),\hdots,f_{n}(\mathbf{c}))^{T}$ with
\begin{gather}
\label{eq:1D_ode_gf}
\widetilde{g}_{i}(\mathbf{c}) = \frac{1}{V_{i}h}\sum_{k\in S_{i}}\overline{D}(c_{i},c_{k})(c_{k}-c_{i}),\quad \widetilde{f}_{i}(\mathbf{c}) = R(c_{i}).
\end{gather}
Recall that our desire here is that the Jacobian of $\widetilde{\mathbf{g}}(\mathbf{c})$, which we denote by $\widetilde{\mathbf{A}}(\mathbf{c})$, is a column transition-rate matrix (Definition \ref{def:1}(vi)). With this in mind, consider the $(i,j)$ entry of $\widetilde{\mathbf{A}}(\mathbf{c})$:
\begin{gather*}
[\widetilde{\mathbf{A}}(\mathbf{c})]_{ij} = \frac{\partial\widetilde{g}_{i}}{\partial c_{j}} = \begin{cases} \frac{1}{V_{i}h}\sum_{k\in S_{i}}[\overline{D}_{c_{i}}(c_{i},c_{k})(c_{k}-c_{i}) - \overline{D}(c_{i},c_{k})], & \text{if $j=i$},\\ \frac{1}{V_{i}h}[\overline{D}_{c_{j}}(c_{i},c_{j})(c_{j}-c_{i}) + \overline{D}(c_{i},c_{j})], & \text{if $j\in S_{i}$},\\ 0, & \text{otherwise}.\end{cases}
\end{gather*}
A standard choice for the inter-nodal diffusivity is the arithmetic mean, $\overline{D}(a,b) = (D(a)+D(b))/2$ (e.g. \cite{simpson_2024,cai_2006}), where $\overline{D}_{a}(a,b) = D^{\prime}(a)/2$ and $\overline{D}_{b}(a,b) = D^{\prime}(b)/2$, which yields
\begin{gather*}
[\widetilde{\mathbf{A}}(\mathbf{c})]_{ij} = \frac{\partial\widetilde{g}_{i}}{\partial c_{j}} = \begin{cases} \frac{1}{2V_{i}h}\sum_{k\in S_{i}}[D'(c_{i})(c_{k}-c_{i}) - D(c_{i})-D(c_{k})], & \text{if $j=i$},\\ \frac{1}{2V_{i}h}[D'(c_{j})(c_{j}-c_{i}) + D(c_{i}) + D(c_{j})], & \text{if $j\in S_{i}$},\\ 0, & \text{otherwise}.\end{cases}
\end{gather*}
Here we see that while $\widetilde{\mathbf{A}}(\mathbf{c})$ is \textit{essentially nonnegative} (Definition \ref{def:1}(iv)) for some choices of $D(c)$, it is not essentially nonnegative in general. For the purpose of this paper, a better choice for the inter-nodal diffusivity is the integrated mean \cite{szymkiewicz_2009}:
\begin{gather}
\label{eq:integrated_mean}
\overline{D}(a,b) = \begin{cases} \frac{1}{b-a}\int_{a}^{b} D(c)\,\text{d}c, & \text{if $a\neq b$,}\\ D(a), & \text{if $a=b$}.\end{cases}
\end{gather}
In this case, $\overline{D}_{a}(a,b) = \frac{1}{(b-a)^{2}}\int_{a}^{b}D(c)\,\text{d}c - \frac{1}{b-a}D(a)$ ($a\neq b$) and $\overline{D}_{b}(a,b) = -\frac{1}{(b-a)^{2}}\int_{a}^{b}D(c)\,\text{d}c + \frac{1}{b-a}D(b)$ ($a\neq b$), and the $(i,j)$ entry of $\widetilde{\mathbf{A}}(\mathbf{c})$ becomes
\begin{gather}
\label{eq:Jacobian_c}
[\widetilde{\mathbf{A}}(\mathbf{c})]_{ij} = \frac{\partial\widetilde{g}_{i}}{\partial c_{j}} = \begin{cases} -\frac{1}{V_{i}h}D(c_{i})|S_{i}|, & \text{if $j=i$},\\ \frac{1}{V_{i}h}D(c_{j}), & \text{if $j\in S_{i}$},\\ 0, & \text{otherwise},\end{cases}
\end{gather}
where $|S_{i}|$ to be the cardinality of $S_{i}$, yielding a Jacobian matrix that is now clearly essentially nonnegative for general nonnegative $D(c)$. While this matrix is now essentially nonnegative it is not, however, a \textit{column transition-rate matrix} (Definition \ref{def:1}(vi)) since each column does not sum to zero due to the presence of the row-dependent control volume length $V_{i}$. However, this can be overcome through a simple change of variables, $c_{i}=V_{i}^{-1}u_{i}$ or equivalently $\mathbf{c}(t) = \mathbf{V}^{-1}\mathbf{u}(t)$ where $\mathbf{V} = \text{diag}(V_{1},\hdots,V_{n})$ and $\mathbf{V}^{-1} = \text{diag}(V_{1}^{-1},\hdots,V_{n}^{-1})$, which produces the ODE system (\ref{eq:ode_nonlinear}) with $\smash{\mathbf{g}(\mathbf{u}) = \mathbf{V}\widetilde{\mathbf{g}}(\mathbf{V}^{-1}\mathbf{u})}$, $\smash{\mathbf{f}(\mathbf{u}) = \mathbf{V}\widetilde{\mathbf{f}}(\mathbf{V}^{-1}\mathbf{u})}$ and $\mathbf{A}(\mathbf{u})$ denoting the Jacobian matrix of $\mathbf{g}(\mathbf{u})$. With $\mathbf{A}(\mathbf{u}) = \mathbf{V}\widetilde{\mathbf{A}}(\mathbf{V}^{-1}\mathbf{u})\mathbf{V}^{-1}$ or $g_{i}(\mathbf{u}) = V_{i}\widetilde{g}_{i}(\mathbf{V}^{-1}\mathbf{u})$, we see that the $(i,j)$ entry of $\mathbf{A}(\mathbf{u})$ is given by
\begin{gather}
\label{eq:Jacobian_u}
[\mathbf{A}(\mathbf{u})]_{ij} = \frac{\partial g_{i}}{\partial u_{j}} = V_{i}\frac{\partial\widetilde{g}_{i}}{\partial c_{j}}V_{j}^{-1} = \begin{cases} -\frac{D(V_{i}^{-1}u_{i})}{V_{j}h}|S_{i}|, & \text{if $j=i$},\\ \frac{D(V_{j}^{-1}u_{j})}{V_{j}h}, & \text{if $j\in S_{i}$},\\ 0, & \text{otherwise}.\end{cases}
\end{gather}
Equivalently, the entries can be expressed column-wise (noting that $j\in S_{i}$ implies $i\in S_{j}$ and vice versa)
\begin{gather}
\label{eq:Jacobian_u}
[\mathbf{A}(\mathbf{u})]_{ij} = \begin{cases} -\frac{D(V_{j}^{-1}u_{j})}{V_{j}h}|S_{j}|, & \text{if $i=j$},\\ \frac{D(V_{j}^{-1}u_{j})}{V_{j}h}, & \text{if $i\in S_{j}$},\\ 0, & \text{otherwise},\end{cases}
\end{gather}
from which we now see that $\sum_{i=1}^{n}[\mathbf{A}(\mathbf{u})]_{ij} = [\mathbf{A}(\mathbf{u})]_{jj} + \sum_{i\in S_{j}}[\mathbf{A}(\mathbf{u})]_{ij} = 0$, for all $j = 1,\hdots,n$. In summary, we have derived a spatial discretisation of the nonlinear reaction-diffusion model (\ref{eq:1D_pde})--(\ref{eq:1D_IBCs})  in the required form of (\ref{eq:ode_nonlinear}) and satisfying the condition that $\mathbf{A}(\mathbf{u})$ is a column transition-rate matrix.
 
\begin{remark}\label{rem:1}\normalfont
Consider the alternative form of the reaction-diffusion model (\ref{eq:1D_pde})--(\ref{eq:1D_IBCs})
\begin{gather}
\label{eq:1D_pde_phi}
\frac{\partial c}{\partial t} = \frac{\partial^{2}\phi}{\partial x^{2}} + R(c),\\
\label{eq:1D_IBCs_phi}
c(x,0) = f(x),\quad \frac{\partial\phi}{\partial x}(0,t) = 0,\quad\frac{\partial\phi}{\partial x}(L,t) = 0.
\end{gather}
Here $\phi$ is the \textit{flux potential} (e.g., \cite{szymkiewicz_2009}) satisfying $\smash{\phi^{\prime}(c) = D(c)}$ with $\smash{\frac{\partial\phi}{\partial x} = \phi^{\prime}(c)\frac{\partial c}{\partial x} = D(c)\frac{\partial c}{\partial x}}$ demonstrating equivalency with the original model (\ref{eq:1D_pde})--(\ref{eq:1D_IBCs}). Discretising (\ref{eq:1D_pde_phi})--(\ref{eq:1D_IBCs_phi}) in space using the same finite volume method as above also yields the ODE system (\ref{eq:1D_ode_c})--(\ref{eq:1D_ode_gf})  with the only difference being that $\widetilde{g}_{i}(\mathbf{c}) = \frac{1}{V_{i}h}\sum_{k\in S_{i}} (\phi(c_{k})-\phi(c_{i}))$. Note, however, that these two different forms of $\widetilde{g}_{i}(\mathbf{c})$ are also equivalent under the integrated mean (\ref{eq:integrated_mean}) since $\smash{\phi(c_{k})-\phi(c_{i}) = \int_{c_{i}}^{c_{k}} \phi'(c)\,\text{d}c = \int_{c_{i}}^{c_{k}}D(c)\,\text{d}c}$. Similarly, the Jacobian matrices remain identical to those given in equations (\ref{eq:Jacobian_c})--(\ref{eq:Jacobian_u}) since $\smash{\phi^{\prime}(c_{i}) = D(c_{i})}$. Hence, the same result can be obtained in a straightforward way by considering (\ref{eq:1D_pde_phi})--(\ref{eq:1D_IBCs_phi}) from the outset. As this equivalency also holds in higher dimensions, this is the approach taken in the next section in 2D.
\end{remark}

\section{Two dimensional formulation}
\label{sec:2D}
Consider the two dimensional analogue of (\ref{eq:1D_pde_phi})--(\ref{eq:1D_IBCs_phi}):
\begin{gather}
\label{eq:2D_pde}
\frac{\partial c}{\partial t} = \nabla^{2}\phi + R(c),\\
\label{eq:2D_IBCs}
c(x,y,0) = F(x,y),\quad \frac{\partial\phi}{\partial x}(0,y,t) = \frac{\partial\phi}{\partial x}(L,y,t) = 0,\quad \frac{\partial\phi}{\partial y}(x,0,t) = \frac{\partial\phi}{\partial y}(x,L,t) = 0,
\end{gather}
where $\phi'(c) = D(c)$. As in 1D, discretisation in space is carried out on a uniform grid consisting of nodes, $(\widehat{x}_{i},\widehat{y}_{j})$ for $i = 1,\hdots,m$ and $j = 1,\hdots,m$, where $\widehat{x}_{i} = (i-1)h$, $\widehat{y}_{j} = (j-1)h$ and $h = L/(m-1)$. The $n = m^{2}$ nodes are ordered left-to-right bottom-to-top such that node $i$ is located at position $x =x_{i} = (p(i)-1)h$ and $y = y_{i}= (q(i)-1)h$ where $p(i) = i - \lfloor\frac{i-1}{m}\rfloor m$ and $q(i) = \lfloor\frac{i-1}{m}\rfloor + 1$. The corresponding finite volumes are then defined by the square regions $[x_{i}^{-},x_{i}^{+}]\times[y_{i}^{-},y_{i}^{+}]$ for $i=1,\hdots,n$, where $x_{i}^{-} = \max(0,x_{i}-h/2)$, $x_{i}^{+} = \min(x_{i}+h/2,L)$, $y_{i}^{-} = \max(0,y_{i}-h/2)$ and $y_{i}^{+} = \min(y_{i}+h/2,L)$. Applying a standard finite volume method by integrating the governing PDE (\ref{eq:2D_pde}) over the finite volumes, inserting the boundary conditions, and employing standard central difference approximations for the diffusion term yields the ODE system:
\begin{gather}
\frac{\text{d}c_{i}}{dt} = \frac{1}{V_{i}}\biggl[\sum_{k\in S_{i}}\alpha_{i,k}\phi(c_{k}) - \phi(c_{i})\sum_{k\in S_{i}}\alpha_{i,k}\biggr] + R(c_{i}),\qquad i = 1,\hdots, n,
\end{gather}
with initial condition $c_{i}(0) = F(x_{i},y_{i})$. Here, $c_{i}\approx c(x_{i},y_{i},t)$, $V_{i} = (x_{i}^{+}-x_{i}^{-})(y_{i}^{+}-y_{i}^{-})$, $S_{i} = \{i-1 \, |\,x_{i} > 0\}\cup\{i+1 \, |\,x_{i} < L\}\cup\{i-m \, |\,y_{i} > 0\}\cup\{i+m \, |\,y_{i} < L\}$ is the set of nodes connected to node $i$, and $\alpha_{i,j} = 1/2$ if both nodes $i$ and $j$ lie on the domain boundary (i.e. ($x_{i}\in\{0,L\}$ or $y_{i}\in\{0,L\}$) and ($x_{j}\in\{0,L\}$ or $y_{j}\in\{0,L\}$), or $\alpha_{i,j} = 1$ otherwise. Equivalently in vector form:
\begin{gather*}
\frac{\text{d}\mathbf{c}}{\text{d}t} = \widetilde{\mathbf{g}}(\mathbf{c}) + \widetilde{\mathbf{f}}(\mathbf{c}),
\end{gather*}
with initial condition $\mathbf{c}(0) = \mathbf{c}_{0}$, where $\mathbf{c} = (c_{1},\hdots,c_{n})^{T}$, $\mathbf{c}_{0} = (F(x_{1},y_{1}),\hdots,F(x_{n},y_{n}))^{T}$; and $\mathbf{g}(\mathbf{c}) = (g_{1}(\mathbf{c}),\hdots,g_{n}(\mathbf{c}))^{T}$ and $\mathbf{f}(\mathbf{c}) = (f_{1}(\mathbf{c}),\hdots,f_{n}(\mathbf{c}))^{T}$ with
\begin{gather*}
\widetilde{g}_{i}(\mathbf{c}) = \frac{1}{V_{i}}\biggl[\sum_{k\in S_{i}}\alpha_{i,k}\phi(c_{k}) - \phi(c_{i})\sum_{k\in S_{i}}\alpha_{i,k}\biggr],\quad \widetilde{f}_{i}(\mathbf{c}) = R(c_{i}).
\end{gather*}
with Jacobian matrix:
\begin{gather*}
[\widetilde{\mathbf{A}}(\mathbf{c})]_{ij} = \begin{cases} -\frac{D(c_{j})}{V_{i}}\sum_{k\in S_{i}}\alpha_{i,k}, & \text{if $j=i$},\\ \frac{D(c_{j})}{V_{i}}\alpha_{i,j}, & \text{if $j\in S_{i}$}\\ 0, & \text{otherwise}.\end{cases}
\end{gather*}
We now apply the same change of variables as in 1D, that is, $c_{i}=V_{i}^{-1}u_{i}$ or equivalently $\mathbf{c}(t) = \mathbf{V}^{-1}\mathbf{u}(t)$ where again $\mathbf{V} = \text{diag}(V_{1},\hdots,V_{n})$ and $\mathbf{V}^{-1} = \text{diag}(V_{1}^{-1},\hdots,V_{n}^{-1})$. Ultimately, this procedure produces the ODE system (\ref{eq:ode_nonlinear}) where $\smash{\mathbf{g}(\mathbf{u}) = \mathbf{V}\widetilde{\mathbf{g}}(\mathbf{V}^{-1}\mathbf{u})}$, $\smash{\mathbf{f}(\mathbf{u}) = \mathbf{V}\widetilde{\mathbf{f}}(\mathbf{V}^{-1}\mathbf{u})}$ and the Jacobian matrix $\mathbf{A}(\mathbf{u})$ is a column transition-rate matrix that can be expressed column-wise by
\begin{gather*}
[\mathbf{A}(\mathbf{u})]_{ij} = \begin{cases} -\frac{D(V_{j}^{-1}u_{j})}{V_{j}}\sum_{k\in S_{j}}\alpha_{k,j}, & \text{if $i=j$},\\ \frac{D(V_{j}^{-1}u_{j})}{V_{j}}\alpha_{i,j}, & \text{if $i\in S_{j}$}\\ 0, & \text{otherwise},\end{cases}
\end{gather*}
with zero column sums since $\sum_{i=1}^{n} [\mathbf{A}(\mathbf{u})]_{ij} = [\mathbf{A}(\mathbf{u})]_{jj} + \sum_{k\in S_{j}} [\mathbf{A}(\mathbf{u})]_{kj} = 0$. In summary, we have derived a spatial discretisation of the 2D nonlinear reaction-diffusion model (\ref{eq:2D_pde})--(\ref{eq:2D_IBCs}) in the required form of (\ref{eq:ode_nonlinear}) and satisfying the condition that $\mathbf{A}(\mathbf{u})$ is a column transition-rate matrix.

\begin{remark}\label{rem:2}\normalfont
As mentioned in the introduction, the standard way to generate stochastic simulations from discretised reaction-diffusion equations is to use a forward Euler discretisation in time. Such a discretisation can be obtained from the exponential integrator (\ref{eq:exponential_integrator}) by approximating $e^{\tau\mathbf{A}_{k}}\approx\mathbf{I} + \tau\mathbf{A}_{k}$ by the first two terms of its power series and $\varphi(\tau\mathbf{A}_{k})\approx \mathbf{I}$ by the first term of its power series, yielding 
\begin{gather}
\label{eq:forward_euler}
\mathbf{u}_{k+1} = (\mathbf{I}+\tau\mathbf{A}_{k})\mathbf{u}_{k} + \tau\mathbf{b}(\mathbf{u}_{k}),
\end{gather} 
which we note is equivalent to applying the forward Euler method directly to the ODE system (\ref{eq:ode_nonlinear}) to give $\mathbf{u}_{k+1} = \mathbf{u}_{k}+\tau\mathbf{g}(\mathbf{u}_{k}) + \tau\mathbf{f}(\mathbf{u}_{k})$ and then adding and subtracting $\tau\mathbf{A}_{k}$. Following the same strategy as outlined in Section \ref{sec:stochastic_framework}, the time-stepping formula (\ref{eq:forward_euler}) can also be used to generate stochastic realisations of $\mathbf{u}_{k}$, denoted by $\widehat{\mathbf{u}}_{k}$ ($k=1,\hdots,M$), where $[\widehat{\mathbf{u}}_{k}]_{j}$ contributes $1/N([\widehat{\mathbf{u}}_{k}]_{j})$ of its value to $[\widehat{\mathbf{u}}_{k+1}]_{i}$ with probability $[\mathbf{I}+\tau\mathbf{A}_{k}]_{ij}$, $[\tau\mathbf{b}(\widehat{\mathbf{u}}_{k})]_{j}$ contributes $1/N(\tau\mathbf{b}(\widehat{\mathbf{u}}_{k})]_{j})$ of its value to $[\widehat{\mathbf{u}}_{k+1}]_{i}$ with probability one and $N(\cdot)$ is defined as in Section \ref{sec:stochastic_framework}. Of course, the requirement here is that $\mathbf{I}+\tau\mathbf{A}_{k}$ is a \textit{column stochastic matrix} (Definition \ref{def:1}(vii)), which is true if $\mathbf{A}_{k}$ is a column transition-rate matrix and $\tau \leq [\mathbf{A}_{k}]_{ij}^{-1}$ for all $i = 1,\hdots,n$ and $j = 1,\hdots,n$. This constraint ensures each probability is in $[0,1]$ (each column automatically sums to one if each column of $\mathbf{A}_{k}$ sums to zero) and simplifies to $\tau \leq h^{2}/(2D_{\text{max}})$ for the 1D formulation (Section \ref{sec:1D}) and $\tau \leq h^{2}/(4D_{\text{max}})$ for the 2D formulation (Section \ref{sec:2D}), where $D_{\text{max}}$ is the maximum value of $D(c)$ attained during the simulation.
\end{remark}

\section{Examples}
\begin{figure}[p]
\includegraphics[width=1.0\textwidth]{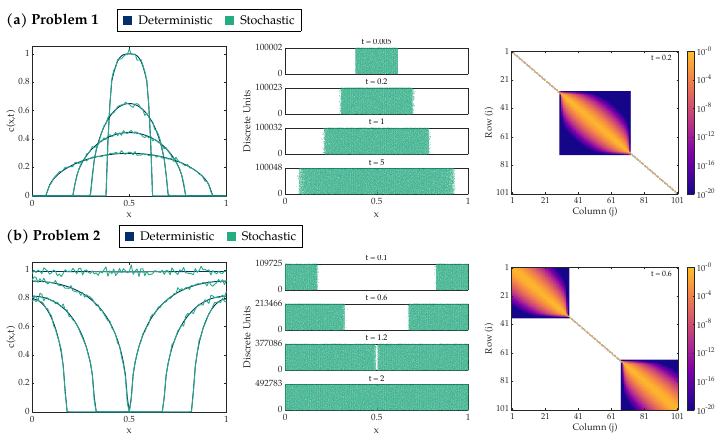}
\caption{\textbf{[1D]}~\textit{First column}: Deterministic solution [plot of $(x_{i},[\mathbf{c}_{k}]_{i})$] compared to one realisation of the stochastic model [plot of $(x_{i},[\widehat{\mathbf{c}}_{k}]_{i})$] at time $t_{k}$ ($t = k\tau$) for $i = 1,\hdots,n$ and $k = k_{1},\hdots,k_{4}$, where $\mathbf{c}_{k} = \mathbf{V}^{-1}\mathbf{u}_{k}$ and $\widehat{\mathbf{c}}_{k} = \mathbf{V}^{-1}\widehat{\mathbf{u}}_{k}$ with $\mathbf{u}_{k}$ and $\widehat{\mathbf{u}}_{k}$ obtained from equation (\ref{eq:exponential_integrator}) and Algorithm \ref{alg:stochastic_model}, respectively. \textit{Second column}: Spatial distribution of discrete units where the number of discrete units located in control volume $i$ at time $t_{k}$ is given by $N([\widehat{\mathbf{u}}_{k}]_{i})$ and the upper limit on the vertical axes is the total number of discrete units in the system i.e. $\sum_{i=1}^{n}N([\widehat{\mathbf{u}}_{k}]_{i})$. \textit{Third column}: Entries of $[\varphi(\tau\mathbf{A}_{k})]_{ij}$ at $k = k_{2}$ ($t = k_{2}\tau$) with white representing a zero value. Parameters: $L = 1$, $n = 101$, $h = 0.01$, $N_{s}=5\times 10^{5}$ (Problem 1 and 2); $T = 5$, $M=1000$, $\tau = 0.005$, $[k_{1},k_{2},k_{3},k_{4}] = [1,40,200,1000]$ (Problem 1); $T=2$, $M=400$, $\tau = 0.005$, $[k_{1},k_{2},k_{3},k_{4}] = [20,120,240,400]$ (Problem~2).}
\label{fig:plot1}
%\end{figure}
%
%\begin{figure}[p]
\medskip
\includegraphics[width=1.0\textwidth]{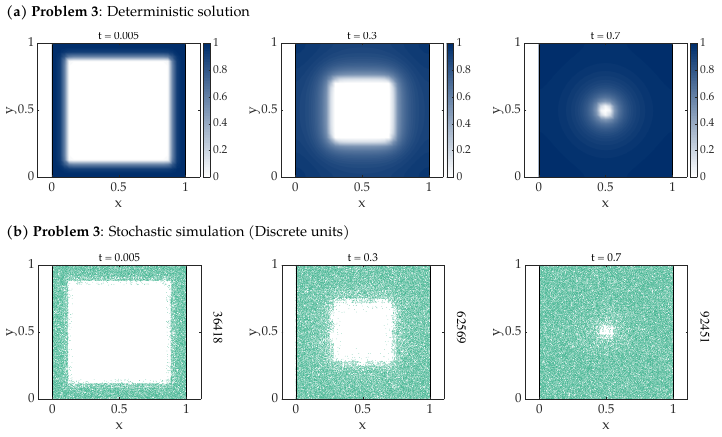}
\caption{\textbf{[2D]} (a) Deterministic solution [plot of $(x_{i},y_{i},[\mathbf{c}_{k}]_{i})$ at $t = k\tau$ for $i = 1,\hdots,n$ and $k = k_{1},k_{2},k_{3}$] where $\mathbf{c}_{k} = \mathbf{V}^{-1}\mathbf{u}_{k}$ with $\mathbf{u}_{k}$ obtained from equation (\ref{eq:exponential_integrator}). (b) Spatial distribution of discrete units at $t = k\tau$ for $k = k_{1},k_{2},k_{3}$, where the number of discrete units located in control volume $i$ at time $t_{k}$ is given by $N([\widehat{\mathbf{u}}_{k}]_{i})$ with $\widehat{\mathbf{u}}_{k}$ obtained from Algorithm \ref{alg:stochastic_model}. The number on the right of the axes is the total number of discrete units in the system i.e. $\sum_{i=1}^{n}N([\widehat{\mathbf{u}}_{k}]_{i})$. Parameters: $L = 1$, $m = 31$, $n = 31^{2}$, $h = 1/30$, $N_{s}=10^{5}$, $T = 1$, $M=200$, $\tau = 0.005$, $[k_{1},k_{2},k_{3}] = [1,60,140]$.}
\label{fig:plot2}
\end{figure}

We now briefly demonstrate the new methodology using the following three test problems: 
\begin{itemize}
\item Problem 1: 1D Porous medium model (\ref{eq:1D_pde})--(\ref{eq:1D_IBCs}) with $D(c) = D_{0}c^{m}$ and $R(c) = 0$, where $D_{0} = 0.1$, $m = 2$ and $F(x) = H(x-0.4)-H(x-0.6)$.
\item Problem 2: 1D Porous Fisher model (\ref{eq:1D_pde})--(\ref{eq:1D_IBCs}) with $D(c) = D_{0}c^{m}$ and $R(c)=\lambda(1-c)c$, where $D_{0} = 0.1$, $m = 2$, $\lambda = 4$ and $F(x) = 1-[H(x-0.1)-H(x-0.9)]$.
\item Problem 3: 2D Porous Fisher model (\ref{eq:2D_pde})--(\ref{eq:2D_IBCs}) with $D(c) = D_{0}c^{m}$ and $R(c)=\lambda(1-c)c$, where $D_{0} = 0.1$, $m = 2$, $\lambda = 9$ and $F(x) = 1-[H(x-0.1)-H(x-0.9)][H(y-0.1)-H(y-0.9)]$.
\end{itemize}
In each problem, $H(\cdot)$ denotes the Heaviside function and the function $\phi$ appearing in equations (\ref{eq:1D_pde_phi})--(\ref{eq:1D_IBCs_phi}) and (\ref{eq:2D_pde})--(\ref{eq:2D_IBCs}) is defined by $\phi(c) = \frac{D_{0}}{m+1}c^{m+1}$ . To compute $e^{\tau\mathbf{A}_{k}}$ and $\varphi(\tau\mathbf{A}_{k})$ we use MATLAB's \texttt{expm} function combined with the identity \cite[Eq (10.56)]{higham_2008}
\begin{gather*}
\mathbf{B} = \begin{bmatrix} \tau\mathbf{A}_{k} &  \mathbf{I}\\ 
\mathbf{0} & \mathbf{0}\end{bmatrix},\qquad 
e^{\mathbf{B}} = \begin{bmatrix} e^{\tau\mathbf{A}_{k}} &  \varphi(\tau\mathbf{A}_{k})\\ 
\mathbf{0} & \mathbf{I}\end{bmatrix},
\end{gather*}
where $\mathbf{0}$ is the zero matrix and $\mathbf{I}$ is the identity matrix. For all three problems, stochastic realisations $\widehat{\mathbf{u}}_{k}$ can be simulated using Algorithm \ref{alg:stochastic_model}, transformed back to the original variable via $\widehat{\mathbf{c}}_{k} = \mathbf{V}^{-1}\widehat{\mathbf{u}}_{k}$ and then compared to the corresponding deterministic solution $\mathbf{c}_{k}$ at chosen discrete times. Results in Figures \ref{fig:plot1} and \ref{fig:plot2} perform this comparison using one realisation of the stochastic model and depict the corresponding positions of the individual discrete units over time. In all problems, we use a time step of $\tau = 0.005$, which exceeds the maximum allowable time step for the corresponding stochastic model obtained using the forward Euler method, where $\tau \leq 0.0005$ (Problems 1 and 2) and $\tau\leq 1/360 \approx 0.0028$ (Problem 3) is required to obtain valid probabilities (see Remark \ref{rem:2} with $D_{\text{max}}=D_{0}=0.1$). For each problem, the stochastic realisation closely mirrors the behaviour of the corresponding deterministic solution with clear sharp fronts of particles invading regions of zero population density. This ability of the stochastic model to capture such regions of compact support becomes evident from the structure of the stochastic matrices $e^{\tau\mathbf{A}_{k}}$ and $\varphi(\tau\mathbf{A}_{k})$, as shown by the examples in Figure \ref{fig:plot1}. This structure stems from the underlying structure of the Jacobian matrix $\mathbf{A}_{k}$, which exhibits a block-diagonal form with zero matrix blocks along the block diagonal corresponding to regions of zero population density. In Problem 1, for example, the Jacobian matrix possesses the structure $\smash{\mathbf{A}_{k} = \text{diag}(\mathbf{0},\mathbf{D}_{k},\mathbf{0})}$ up until the point in time when the left and right moving fronts reach the boundary of the domain ($x = 0,L$) with the size of $\mathbf{D}_{k}$ increasing up until this point. In this case, the identity for a matrix function applied to a block-diagonal matrix \cite[Thm 1.13(g)]{higham_2008} yields $\smash{e^{\tau\mathbf{A}_{k}} =\text{diag}(\mathbf{I},e^{\tau\mathbf{D}_{k}},\mathbf{I})}$ and $\smash{\varphi(\tau\mathbf{A}_{k}) =\text{diag}(\mathbf{I},\varphi(\tau\mathbf{D}_{k}),\mathbf{I})}$, with the off-diagonal zero matrix blocks and diagonal identity matrix blocks making it impossible for transitions to occur between regions of zero and nonzero population density beyond the invading front.

\subsection*{Data availability}
Supporting MATLAB code will be made available on GitHub once the paper is published.

\newpage
\begingroup
\setlength{\bibsep}{0pt}
\linespread{0.91}\selectfont
\bibliographystyle{unsrt}
\bibliography{references}
\endgroup

\end{document}